\documentclass[aps,prl,nofootinbib,superscriptaddress,twocolumn]{revtex4}
\usepackage{graphicx}
\usepackage{color}

\newcommand{\bnmr}{$\beta$-NMR}

\newcommand{\sto}{SrTiO$_3$}
\newcommand{\lao}{LaAlO$_3$}

\newcommand{\Li}{$^{8}$Li}

\begin{document}

\title{Nature of weak magnetism in SrTiO$_3$/LaAlO$_3$ multilayers}

\author{Z.~Salman}
\email{zaher.salman@psi.ch}
\affiliation{Laboratory for Muon Spin Spectroscopy, Paul Scherrer Institut, CH-5232 Villigen PSI, Switzerland}
\author{O.~Ofer}
\affiliation{TRIUMF, 4004 Wesbrook Mall, Vancouver, BC, Canada, V6T 2A3}
\author{M. Radovic}
\affiliation{Swiss Light Source, Paul Scherrer Institut, CH-5232 Villigen PSI, Switzerland}
\affiliation{Laboratory for Synchrotron and Neutron Spectroscopy, Ecole Polytechnique Federale de Lausanne, CH-1015 Lausanne, Switzerland}
\author{H.~Hao}
\affiliation{Department of Physics and Astronomy, University of
  British Columbia, Vancouver, BC, Canada V6T 1Z1}
\author{K.~H.~Chow}
\affiliation{Department of Physics, University of Alberta, Edmonton,
  AB, Canada T6G 2E1}
\author{M.~D.~Hossain}
\affiliation{Department of Physics and Astronomy, University of
  British Columbia, Vancouver, BC, Canada V6T 1Z1}
\author{C.~D.~P.~Levy}
\affiliation{TRIUMF, 4004 Wesbrook Mall, Vancouver, BC, Canada, V6T 2A3}
\author{W.~A.~MacFarlane}
\affiliation{Department of Chemistry, University of British Columbia,
  Vancouver, BC, Canada V6T 1Z1}
\author{G.~M.~Morris}
\affiliation{TRIUMF, 4004 Wesbrook Mall, Vancouver, BC, Canada, V6T 2A3}
\author{L.~Patthey}
\affiliation{Swiss Light Source, Paul Scherrer Institut, CH-5232 Villigen PSI, Switzerland}
\author{M.~R.~Pearson}
\affiliation{TRIUMF, 4004 Wesbrook Mall, Vancouver, BC, Canada, V6T 2A3}
\author{H.~Saadaoui}
\affiliation{Laboratory for Muon Spin Spectroscopy, Paul Scherrer Institut, CH-5232 Villigen PSI, Switzerland}
\author{T.~Schmitt}
\affiliation{Swiss Light Source, Paul Scherrer Institut, CH-5232 Villigen PSI, Switzerland}
\author{D.~Wang}
\affiliation{Department of Physics and Astronomy, University of
  British Columbia, Vancouver, BC, Canada V6T 1Z1}
\author{R.~F.~Kiefl}
\affiliation{Department of Physics and Astronomy, University of
  British Columbia, Vancouver, BC, Canada V6T 1Z1}
\affiliation{TRIUMF, 4004 Wesbrook Mall, Vancouver, BC, Canada, V6T 2A3}

\begin{abstract}
  We report the observation of weak magnetism in superlattices of
  \lao/\sto\ using $\beta$-detected nuclear magnetic resonance. The
  spin lattice relaxation rate of \Li\ in superlattices with a spacer
  layers of 8 and 6 unit cells of \lao\ exhibits a strong peak near
  $\sim 35$~K, whereas no such peak is observed in a superlattice with
  spacer layer thickness of 3 unit cells. We attribute the observed
  temperature dependence to slowing down of weakly coupled electronic
  moments at the \lao/\sto\ interface. These results show that the
  magnetism at the interface depends strongly on the thickness of the
  spacer layer, and that a minimal thickness of $\sim 4-6$ unit cells
  is required for the appearance of magnetism. A simple model is used
  to determine that the observed relaxation is due to small
  fluctuating moments ($\sim 0.002$ $\mu_B$) in the two samples with a
  larger \lao\ spacer thickness.
\end{abstract}

\maketitle

The electronic, magnetic and structural properties of an interface
between two materials is in general different from the bulk properties
of both. A dramatic example, discovered recently
\cite{Ohtomo04N,Thiel06S,Huijben06NM}, is the high mobility
two-dimensional electron gas (2DEG) at the interface between two
insulating perovskite oxides; TiO$_2$-terminated SrTiO$_{3}$ (STO) and
LaAlO$_{3}$ (LAO). Surprisingly, there is evidence that this interface
can be both magnetic\cite{Brinkman07NM,BenShalom09PRB} and even
superconducting below $\sim 300$~mK\cite{Reyren07S}. It is generally
agreed that these properties are associated with subtle structural
changes at the interface. Several attempts have been made to explain
the high carrier densities at the interface, including doping with
electrons or oxygen vacancies \cite{Thiel06S, Pentcheva06PRB,
  Park06PRB, Takizawa06PRL, Kalabukhov07PRB}, inter-diffusion
\cite{Takizawa06PRL, Nakagawa06NM, Willmott07PRL}, and the influence
of lattice distortions \cite{Ahn03N, Gemming06AM,
  Hamann06PRB,Maurice06PSS,Okamoto06PRL,Zhou11RPB}. However, the
details and mechanism behind the observed properties seem to involve
several processes
\cite{Hwang06S,Nakagawa06NM,Kalabukhov07PRB,Siemons07PRL,Herranz07PRL,Pentcheva08PRB,Willmott07PRL}.

The unusal properties of the LAO/STO interface is extremely relevant
for the general interface phenomena at oxide and perovskite interfaces
\cite{Dagotto07S,Hwang12NM}. This important class of materials
exhibits a variety of physical properties including magnetic
\cite{Boris11S,Dong08PRB,Kida07PRL,Adamo08APL}, superconducting
\cite{Reyren07S}, insulating and conducting
\cite{Ohtomo02N,Ohtomo04N,Reyren07S,Boris11S}. The observed weak
magnetism in this particular LAO/STO system may have significant
implications on the interpretation of interface properties and
proximity effects in other oxides. However, since both materials in
this case are non-magnetic and insulating the appearance of weak
magnetism and conductivity can be easily detected.

In this Letter we address questions concerning the nature of the
reported magnetism at the interfaces between LAO and STO. To date,
most reports of magnetism at these interfaces are indirect being based
on transport measurements at high applied magnetic field and limited
to bi-layers. More recent reports have contradicting claims of
coexistence \cite{Dikin11PRL,Bert11NP} and phase separation
\cite{Ariando11NC} of superconductivity and magnetism. These studies
report measurements on bi-layers of LAO on TiO$_2$ terminated STO, and
currently more efforts are being invested in producing superlattices
(SLs) of LAO/STO\cite{Zhou11RPB,Fitzsimmons11PRL}. This is to address
the question whether the interfaces in SLs maintain the same
properties as bi-layers, but also to answer the question whether the
TiO$_2$ termination of the STO substrate, and the subsequent polar
ctastrophe scenario, is crucial in this case
\cite{Ohtomo04N,Siemons07PRL}. Recent polarized neutron reflectometry
(PNR) has concluded there is no detectable magnetism in SLs, putting a
very small upper limit on any possible magnetization
\cite{Fitzsimmons11PRL}. In fact, until now there has been no direct
observation of internal magnetic fields (in either bi-layers or SLs)
that must be present in any true magnetic state.  Here we report such
results using $\beta$-detected nuclear magnetic resonance
($\beta$-NMR) measurements in SLs of LAO/STO. For SLs with a LAO
spacer layer exceeding a ``critical'' thickness, we find the spin
lattice relaxation rate of polarized \Li\ exhibits a strong
temperature dependence with a maximum at $T^* \sim 35$~K. This
behaviour is typical of a slowly fluctuating internal magnetic field
expected near a magnetic transition at $T^*$, and provides direct
evidence of magnetism at the interface between insulating and
nonmagnetic LAO and STO. The weak magnetism is attributed to localized
charge carriers at the interface. We estimate that the size of the
magnetic moment per unit cell (uc) is about $\sim 1.8 \times 10^{-3}$
$\mu_B$, indicating the moments are only weakly dependent of the LAO
spacer thickness beyond a critical value of $4-6$~uc.

The \bnmr\ technique is a magnetic resonance technique similar to both
nuclear magnetic resonance and muon spin relaxation ($\mu$SR). The
local spin probe used here is \Li. A low energy ($28$ keV) beam of
radioactive \Li\ is produced at the isotope separator and accelerator
(ISAC) at TRIUMF in Vancouver, Canada. It is then spin-polarized using
a collinear optical pumping method, yielding nuclear polarization in
excess of $70 \%$, and subsequently implanted into the sample. Since
the implanted beam energy can be adjusted, the \Li\ mean stopping
depth can be varied between 1-250 nm. The nuclear polarization, and
its time evolution, is the quantity of interest in these experiments.
It can be measured through the $\beta$-decay asymmetry, where an
electron is emitted preferentially opposite to the direction of the
nuclear polarization at the time of decay \cite{Crane01PRL} and
detected by appropriately positioned scintillation counters. \Li\ is a
spin $I=2$ nucleus with a small electric quadrupole moment $Q=+31$~mB
and gyromagnetic ratio $\gamma=6.301$ MHz/T. The spin lattice
relaxation of the \Li\ nuclear spin can be measured by implanting a
short pulse of beam for a duration $t_p$ (e.g. $1$ second), and
measuring the polarization as a function of time, $p_z(t)$, during and
after the beam pulse.  More details about the techniques can be found
in Refs.~\cite{Salman04PRB,Salman06PRL,Salman07NL}.

Measurements on three different SLs are reported here. These were
grown using pulsed laser deposition and consist of 10 LAO/STO stacking
periods grown on TiO$_2$ terminated $\langle 100 \rangle$ single
crystal STO substrates. The thickness of the LAO layers were $n=8$, 6
and 3 uc, while the STO layers are fixed at 10 uc \cite{Zhou11RPB}.
Hereafter, we refer to these SLs as LAO$n$, where $n$ is the number of
uc in the LAO layers. After growth, the samples were annealed for 5
hours at $1000^\circ$C in 1 bar of O$_2$ in order to fill oxygen
vacancies \cite{Zhou11RPB}. All LAO$n$ samples were investigated using
resonant inelastic X-ray scattering (RIXS), and their preparation
details are given in Ref.~\cite{Zhou11RPB}.  Additional control
measurements were also performed on STO and LAO single crystals
obtained from Crystec GmBH.

\begin{figure}[h]
  \centerline{\includegraphics[width=0.8\columnwidth]{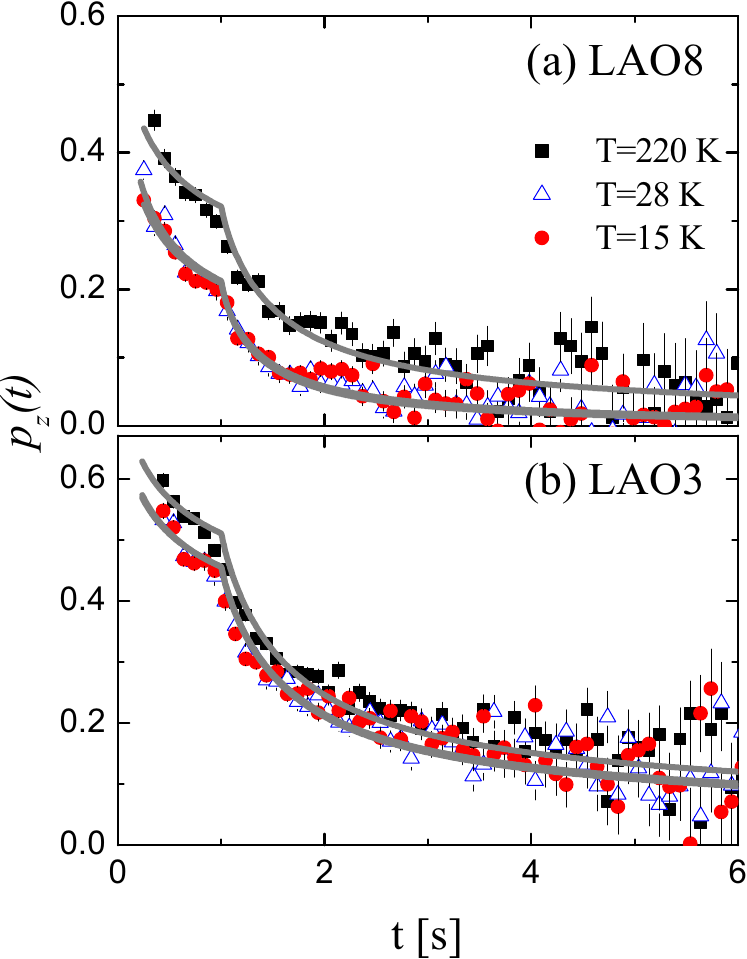}}
  \caption{(Color online) \Li\ spin relaxation curves measured in (a)
    LAO$8$ and (b) LAO$3$ in 3 mT applied field, 5 keV \Li\
    implantation energy and various temperatures. Note the relaxation
    rate is larger and more temperature dependent in LAO$8$ compared
    to LAO$3$. Note also the long lived tail in LAO$3$ which is absent
    in LAO$8$.}
\label{Asy}
\end{figure}

\begin{figure}[h]
  \centerline{\includegraphics[width=0.8\columnwidth]{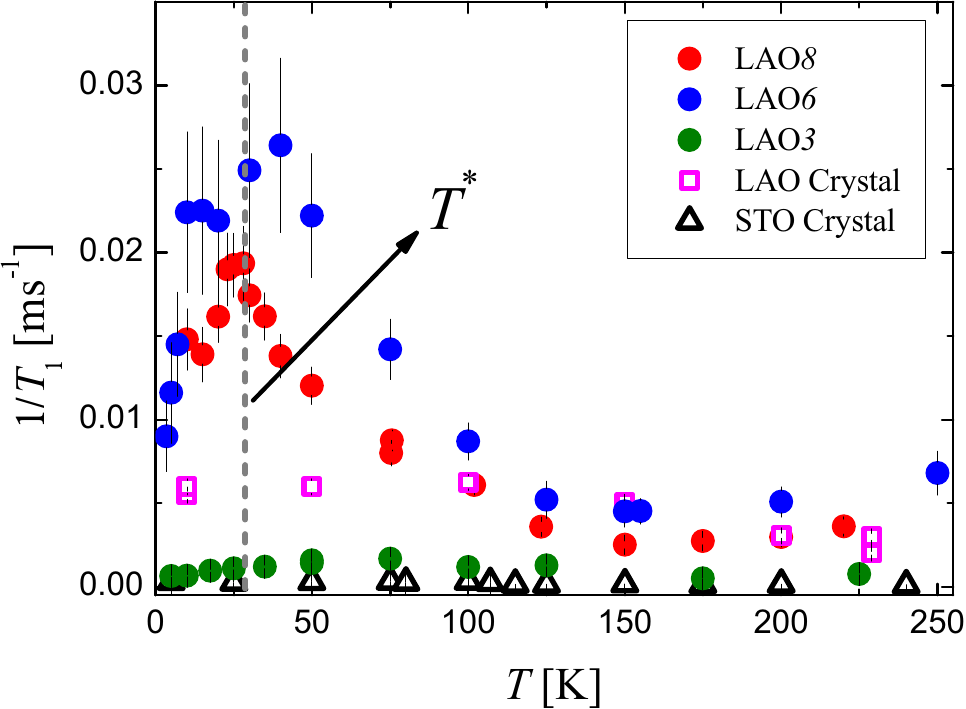}}
  \caption{(Color online) The spin lattice relaxation rate ($1/T_1$)
    as a function of temperature in 3 mT applied field. The red, blue
    and green circles are measurements in LAO$8$, LAO$6$ and LAO$3$,
    respectively. The squares and triangles are reference measurements
    in LAO and STO bare crystals.}
\label{MultiLayer}
\end{figure}

Typical relaxation curves measured in LAO$8$ and LAO$3$, using 5 keV
\Li\ implantation energy, are shown in Fig.~\ref{Asy}(a) and (b),
respectively. This implantation energy corresponds to mean
implantation depth of $\sim 20$ nm in the samples.
In these measurements $p_z(t)$ is determined by both the \Li\
spin-lattice relaxation rate, $\lambda=1/T_1$, and its radioactive
lifetime, $\tau=1.21$s. Assuming a beam pulse duration $t_p$
and a general spin relaxation function $f(t,t':\lambda)$ for the
fraction of \Li\ implanted in the sample at $t'$, the polarization
follows \cite{Salman06PRL}
\begin{equation} \label{genrlx}
p_z(t)=\left\{
\begin{array}{ll}
  \frac{\int_0^t e^{-(t-t')/\tau} f(t,t':\lambda)
    dt'}{\int_0^t e^{-t/\tau} dt} & t \le t_p \\
  \frac{\int_0^{t_p} e^{-(t_p-t')/\tau} f(t,t':\lambda)
    dt'}{\int_0^{t_p} e^{-t/\tau} dt} & t > t_p.
\end{array}
\right.
\end{equation}
The data in Fig.~\ref{Asy} are best fit to Eq.~(\ref{genrlx}) with a
phenomenological stretched-exponential form,
\begin{equation}
f(t,t':\lambda)=Ae^{-\left[ \lambda(t-t') \right]^{0.3}}.
\end{equation}
A much stronger temperature dependence is observed in both LAO$8$ and
LAO$6$, with a relaxation rate which is generally higher than that
observed in LAO$3$. In Fig.~\ref{MultiLayer}
we plot the relaxation rates in all SLs as a function of temperature
compared to the relaxation rates measured in single crystals of LAO
and STO under the same conditions. As expected, the relaxation in STO
is much smaller than that measured in LAO, where the fluctuating Al
nuclear moments contribute to the relaxation of the \Li\ spin at these
low fields \cite{Salman06PRL,Salman07PRB}. Note, in both STO and LAO
single crystals, there is only weak temperature dependence. LAO$3$
exhibits a similar weak temperature dependence, while LAO$8$ and
LAO$6$ show a clear increase in $1/T_1$ as the temperature is lowered,
followed by a pronounced peak near $T^* \sim 35$~K and a decrease as
the temperature is lowered further. Note, the strong temperature
dependence and enhancement in $1/T_1$ at low temperatures in LAO$6$
and LAO$8$ is not present in intrinsic LAO or STO. We attribute this
behaviour to slow magnetic fluctuations, due to magnetic freezing or
critical slowing down of magnetic fluctuations at the interfaces
occurring near $T^*$.

The relaxation in LAO$8$ and LAO$6$ at high temperature approaches
that measured in bulk LAO, while in LAO$3$ the relaxation at high
temperature is somewhere between that of LAO and STO. This difference
can be understood as an average contribution of LAO and STO layers in
the different SLs. At the 5 keV \Li\ implantation energy we estimate
that the ratio between \Li\ stopped in STO:LAO is $\sim 1.3:1$ in
LAO$8$, $\sim 1.7:1$ in LAO$6$, and $\sim 3.4:1$ in LAO$3$.
Therefore, our results indicate that at high temperatures the
relaxation in the SLs is dominated by fluctuating nuclear moments in
LAO (though some contribution of the magnetic fluctuations are still
present in LAO$8$ and LAO$6$). However, at lower temperatures there is
clear evidence of a different relaxation mechanism developing in
LAO$8$ and LAO$6$, which is not present in LAO$3$ or intrinsic STO or
LAO.

Recent RIXS measurements on the same samples have revealed localized
as well as delocalized Ti 3d carriers in such SLs \cite{Zhou11RPB}.
These were attributed to spin-bearing Ti$^{3+}$ ions at the interface.
An orthorhombic structural distortion of Ti$^{3+}$O$_6$ octahedra was
also observed. However, while the density of charge carriers depends
on the thickness of the LAO layers, $n$, the distortion of the
Ti$^{3+}$O$_6$ does not. Furthermore, the annealing process was found
to reduce significantly the density of both types of carriers due to
the reduction of Ti$^{3+}$ to Ti$^{4+}$\cite{Muller04N}, but it does
not affect the orthorhombic distortion at the interfaces. From these
measurements it was concluded that, for the annealed SLs, there is a
critical thickness of 6 LAO uc, above which the density of carriers
increases dramatically \cite{Zhou11RPB}.

Our spin lattice relaxation measurements demonstrate that the LAO$8$
and LAO$6$ samples exhibit significantly enhanced spin relaxation at
low temperatures compared with LAO$3$. More importantly, we see a
distinct anomaly near $T^*$, possibly related to the onset of the
static magnetism reported near 35 K \cite{BenShalom09PRB}. A priori,
the peak at $T^*$ could have a non-magnetic origin. For example,
temperature dependent fluctuations in the electric field gradient
(EFG) at the \Li\ site, which couple to its electric quadrupole moment
\cite{Salman06PRL} (e.g. a ferroelectric transition). However, we can
rule out EFG fluctuations since (I) RIXS measurements confirm that the
non-cubic distortions in these SLs do not depend on the thickness of
the LAO layers (and so do their contributions to $1/T_1$), and (II) we
do not observe a strong temperature dependence in LAO$3$. Hence, the
$1/T_1$ enhancement in LAO$8$ and LAO$6$ must have a magnetic origin,
and therefore, almost certainly due to localized charge carriers at
the interface. In what follows, we evaluate the average size of the
magnetic moments per unit cell, assuming that the magnetism is
concentrated at the LAO/STO interfaces.

The \Li\ probes are implanted almost uniformly within the volume of
the SLs. Using our $1/T_1$ results in the magnetic SLs we can estimate
the size of fluctuating local magnetic fields, $\Delta$, experienced
by the \Li. In the fast fluctuation limit we can write
\cite{Slichter},
\begin{equation} \label{SLR}
\frac{1}{T_1}=\frac{\gamma^2 \Delta^2 \tau_c}{1+\omega^2 \tau_c^2},
\end{equation}
where $\tau_c$ is the correlation time of magnetic field fluctuations
and $\omega$ is the precession frequency of the spin probe. In the
presence of strong quadrupolar interactions, as in STO and LAO,
$\omega$ is dominated by the quadrupolar frequency of the transition
$m=\pm2 \rightarrow \pm1$. This can be estimated at $\sim 230$ kHz in
STO \cite{Salman04PRB}. We assume for simplicity that the maximum in
$1/T_1$ corresponds to a $T_1$ minimum such that $\tau_c$ satisfies
$\omega \tau_c \sim 1$ \cite{Slichter}. In this case, we can estimate
$\Delta \simeq 4.8 \times 10^{-4}$ and $5.4 \times 10^{-4}$ T for the
LAO$8$ and LAO$6$, respectively.

\begin{figure}[h]
  \centerline{\includegraphics[width=0.8\columnwidth]{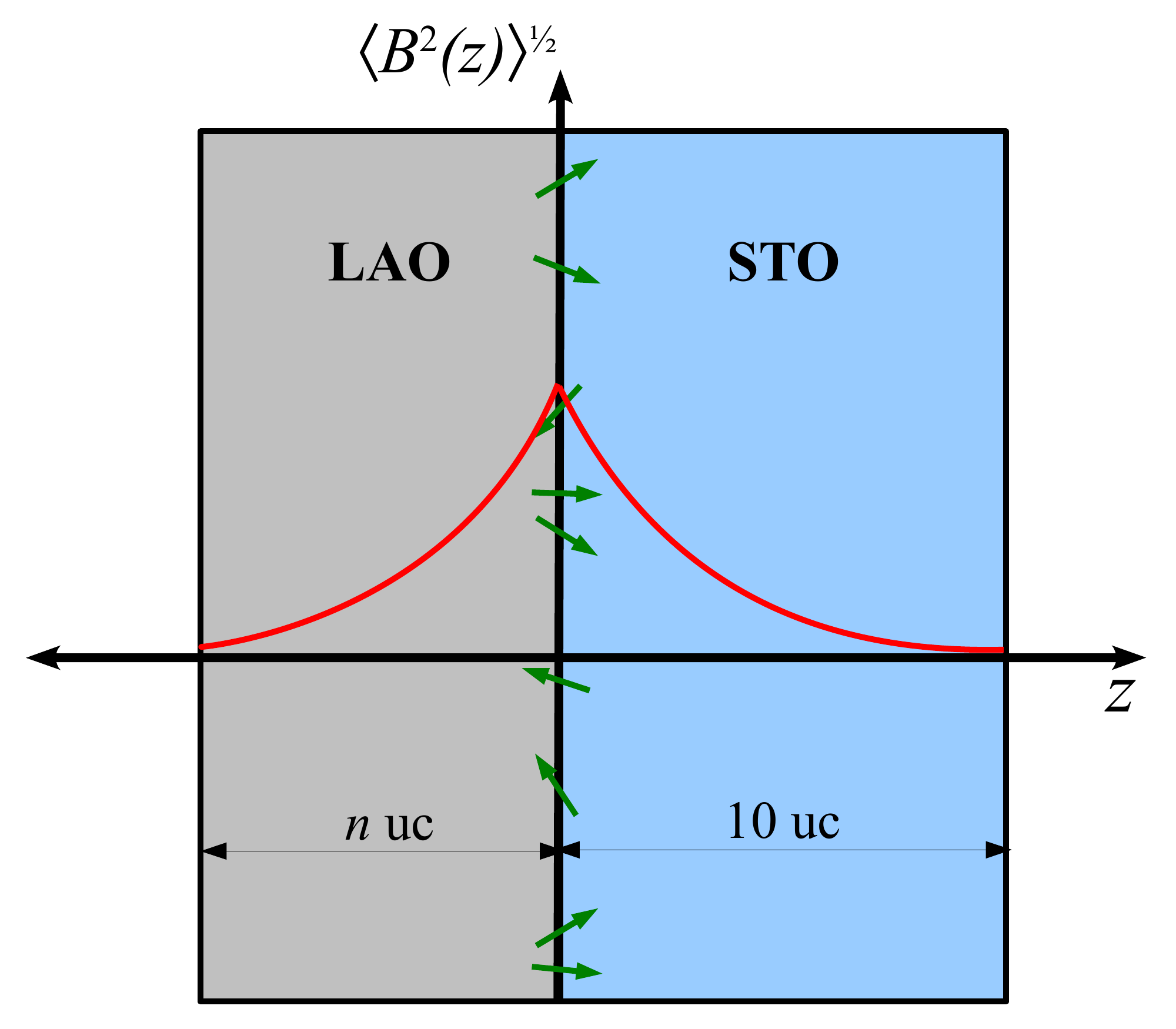}}
  \caption{(Color online) A schematic of magnetic moments (green
    arrows) at the LAO/STO interface. The red lines represent are the
    RMS of dipolar field distribution experienced by the \Li.}
\label{Schematic}
\end{figure}

One can also estimate the size of the moment needed to produce such
magnetic fields using a few simplifying assumptions. First we assume
there is a lattice of magnetic moments, $\mu=\alpha \mu_B$ ($\alpha$
is a constant and $\mu_B$ is the Bohr magneton), arranged on a square
lattice ($a$) at the interfaces. We then calculate the distribution of
dipolar fields experienced by a \Li, located at a distance $z$ from
the interface, by summing up the contributions from all moments
\cite{Salman07NL,Salman12PP} (see schematic in Fig.~\ref{Schematic}).
The root mean square (RMS) of the distribution falls as $\sim 1/z^2$
away from the interface \cite{Salman07NL,Salman12PP}. Therefore, the
resulting RMS averaged over all implanted \Li\ (assuming a uniform
distribution within all layers) is,
\begin{equation} \label{MLCalc}
\Delta_{\rm th} \simeq C_0 \frac{\alpha}{a^3},
\end{equation}
where $C_0$ is a parameter that depends on the LAO layer thickness,
$a$ is in units of \AA, and the resulting $\Delta_{\rm th}$ is in
Tesla. From these calculations we find $C_0=17.36$ for LAO$8$ and
18.95 for LAO$6$. Taking $a \sim 4$ \AA\ as the unit cell of LAO (and
STO) we find that $\Delta_{\rm th} \simeq 0.271 \alpha$ T and $0.296
\alpha$ T for LAO$8$ and LAO$6$, respectively. Note that for
$\mu=1\mu_B$, $\Delta_{\rm th}$ is about two to three orders of
magnitude larger than the $\Delta$ estimated from $1/T_1$. Thus, our
measurements imply an average magnetic moment of $\sim 1.8 \times
10^{-3}$ $\mu_B$ per unit cell at the LAO/STO interfaces, {\em equal
  in both} LAO$8$ and LAO$6$ samples. The difference in $1/T_1$ is
simply due to the different thickness of LAO layers. This result is
consistent with (and confirms) the assumption that in the magnetic SLs
the moments are confined to the interfaces and further indicates that
their average size is independent of the LAO spacer layer thickness
beyond the critical value. The small magnetic moment also explains why
it has been missed with less sensitive techniques such as PNR
\cite{Fitzsimmons11PRL}.

It is likely that the observed magnetization is not uniformly
distributed over the interface. If instead, we assume that there is an
inhomogeneous distribution of $1 \mu_B$ moments, then our calculations
imply a two-dimensional spin density of $\sim 1.13 \times 10^{12}$
$\mu_B/{\rm cm}^{2}$. Surprisingly, this is of the same order of
estimates from scanning SQUID measurements in bi-layers, $\sim
7.3\pm3.4 \times 10^{12}$ $\mu_B/{\rm cm}^{2}$ \cite{Bert11NP}. The
small difference could be simply due to a different sample preparation
procedure or a difference between bi-layers and superlattices.
Moreover, there is a fundamental difference between the two. In a
bi-layer the interface is formed between a TiO$_2$ terminated STO and
a LaO$^+$ terminated layer of LAO, i.e.  TiO$_2$/LaO$^+$. In contrast,
it can be either TiO$_2$/LaO$^+$ or SrO/AlO$_2^-$ in the SLs. These
two types of interfaces have dramatically different electronic
properties \cite{Ohtomo04N}, in relation to the polar catastrophe
\cite{Siemons07PRL} due to the different net charge of the LaO$^+$ and
AlO$_2^-$ layers. It is important to point out here that our results
are consistent with the magnetism residing on {\em both types} of
interfaces. Finally, the broad $1/T_1$ peak (in temperature) in the
magnetic samples is further indication of the dilute and disordered
magnetic moments at these interfaces (typically seen in dilute spin
glasses \cite{Keren96PRL}), in agreement with Ref.~\cite{Bert11NP}.

In conclusion, $\beta$-NMR of low energy \Li\ was used to investigate
SLs of LAO/STO. We present direct evidence for weak magnetism in these
SLs, attributed to a dilute concentration of magnetic moments at the
interfaces. Our measurements agree with previous reports of this
phenomenon in bi-layers of LAO/STO
\cite{Brinkman07NM,BenShalom09PRB,Dikin11PRL,Bert11NP,Ariando11NC},
but exhibit a surprising dependence on the thickness on the LAO
layers. The magnetism is observed only in SLs with LAO layers
exceeding a ``critical'' thickness of $4-6$~uc. This provides strong
evidence for a direct connection between the observed magnetism and
localized charge carriers detected in RIXS \cite{Zhou11RPB}.
Furthermore, we find that the magnetism seems to be highly disordered
and displays evidence of critical slowing down and possibly freezing
near $T^* \sim 35$~K. A simple model calculation shows that it can be
attributed to a two-dimensional spin density of localized magnetic
moments of $\sim 1.13 \times 10^{12}$ $\mu_B/{\rm cm}^{2}$ which is
independent of the thickness of LAO layers in magnetic SLs. This value
is slightly lower than that found in bi-layers \cite{Bert11NP},
nevertheless, it could explain its absence in the PNR data
\cite{Fitzsimmons11PRL}, since it does not produce sufficient contrast
between the opposite neutron polarizations. Furthermore, our results
demonstrate that, unlike the 2DEG, the magnetism apears on both types
of STO/LAO interfaces, and therefore is unrelated to the polar
catastrophe scenario. This indicates that the mechanism behind the
2DEG and magnetism may be different. Finally, our results establish a
very stringent test for any robust theory attempting to explain the
observed phenomena at the LAO/STO interfaces. We also note that these
results may have significant implications on the interpretation of
interface phenomena in oxide and perovsike materials in general
\cite{Dagotto07S,Hwang12NM}.

\begin{acknowledgments}
  This research was supported by the Center for Materials and
  Molecular Research at TRIUMF, NSERC of Canada and the CIFAR. We
  would like to acknowledge Rahim Abasalti, Bassam Hitti, Donald
  Arseneau, and Suzannah Daviel for expert technical support. We are
  grateful for Prof. J. Mesot for reading the manuscript and fruitful
  discussions.
\end{acknowledgments}

\newcommand{\noopsort}[1]{} \newcommand{\printfirst}[2]{#1}
  \newcommand{\singleletter}[1]{#1} \newcommand{\switchargs}[2]{#2#1}

\end{document}